\documentclass[journal=jacsat,manuscript=communication]{achemso}

\usepackage[version=3]{mhchem} 

\usepackage{graphicx}

\usepackage{amsmath}
\usepackage{amssymb}
\usepackage{wasysym}
\usepackage{textcomp}

\author{Vinicius M. Lenart}
\affiliation{Universidade Estadual de Ponta Grossa}
\author{Sergio L. G\'omez}
\affiliation{Universidade Estadual de Ponta Grossa}
\author{Maria P. Calatayud}
\affiliation{Universidad de Zaragoza}
\author{Gerardo R. F. Goya}
\affiliation{Universidad de Zaragoza}
\email{goya@unizar.es}

\title[]{Size and shape control of magnetite nanoparticles with a nonselective binding surfactants}

\abbreviations{TEM, SEM, SAR, FTIR}
\keywords{-}

\begin{document}

\section{Introduction}

A Martian meteorite\cite{martian}, magnetic inks \cite{ink}, drug targeting, batteries, contrasts for MRI, data storage or even clinical thermo-therapy\cite{batteries,medicalapplications} seem to have no connection, but all have in common a dark mineral called magnetite. However, in each of these applications this iron oxide shows up with different forms because their optical, electrical and magnetic properties are strongly dependent on size, shape and kind of surfactant. In this sense the control of this characteristics has long been of scientific and technological interest. In an AC magnetic field-assisted cancer therapy, \textit{e.g.}, from a biological point of view, the interaction of the nanoparticles with cells is critically determined by the surface properties which control their fate in biological environments\cite{surface1}. Besides the size, factors as the shape also seems to affect the cellular uptake  \cite{shapeinfluence1, shapeinfluence2,shapeinfluence4, shapeinfluence5, shapeinfluence6}. On the other hand, the specific absorption rate (SAR) at a fixed frequency and magnetic field, is hugely dependent on average and distribution of size, shape, crystalline anisotropy, and degree of aggregation or agglomeration of the nanoparticles\cite{spa1,spa2}. Each of these factors contributes to an independent energy loss mechanism: N\'eel relaxation, Brown relaxation or magnetic hysteresis loss\cite{hypertermia}. Thus, the key for improving the efficiency of a given application is the knowledge about the morphological control.

\section{Results and discussion}

\begin{figure}[H]
\centering
  \includegraphics[scale=1.1]{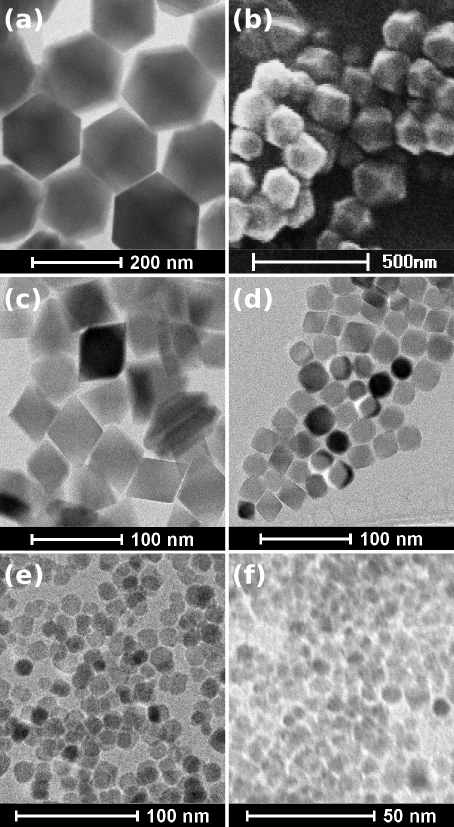}
\caption{TEM (a) and SEM (b) images of   truncated octahedron of Fe$_3$O$_4$ nanoparticles synthesized without oleylamine ($\upsilon=0$), with 176\,nm approximately. Figures (c) and (d) show a mix of octahedron and cube nanoparticles of 63, 29 prepared with 4 and 8\,mmol respectively. Increasing the amount of oleylamine to 12 and 16\,mmol result in spherical-like nanoparticles with 17 (e) and 6.5\,nm (f).}
\label{fig:1}
\end{figure}

\begin{figure}
\centering
\includegraphics[scale=0.3]{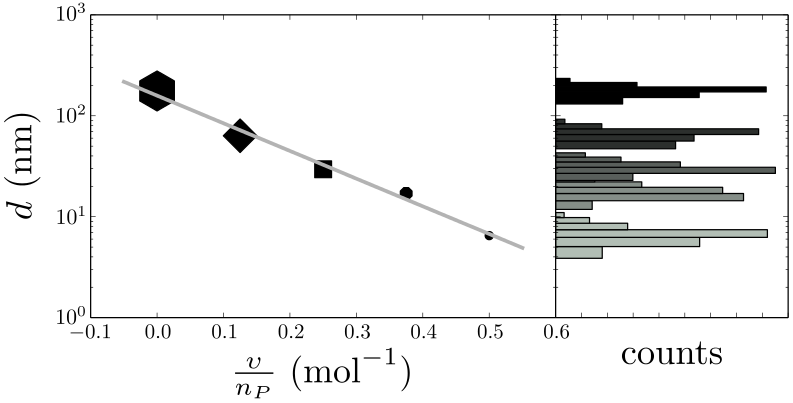}
\caption{Size as a function of the ratio of surfactant and precursor. The surfactant contribution is expressed by shape factor $\upsilon$ defined in equation \ref{eq:1}. The right side shows the distribution of size. The gray line is the best fit and is express by the equation \ref{eq:2}.}
\label{fig:2}
\end{figure}

In this sense we synthesized magnetite nanoparticles \textit{via} typical thermal decomposition method. We used 4\,mmol of Fe(acac)$_3$ as a precursor, 20\,ml of benzyl ether as a solvent and for the first layer of surfactant a mix of 8\,mmol of oleic acid and (0, 4, 8, 12, 16)\,mmol of oleylamine\cite{Sun2002, dokyoon}. Each sample was desoxigened at room temperature for 30 minutes, then heated to 295\,\textcelsius \ at a rate of approximately 10\,\textcelsius/min, keeping at that temperature for 30 minutes and then cooled to room temperature. The entire process was carried out in a nitrogen atmosphere. In the second step the final product of the synthesis was washed several times with hexane and alcohol and dispersed in an organic solvent. Then, the nanoparticles were dispersed in an aqueous medium through the second coating with lauric acid which shows good results of intracellular uptake\cite{lauric}. At the end of this process we obtained ferrofluids in aqueous medium at physiological pH and high concentrations of about 9.9\,mg/ml of Fe$_3$O$_4$. Therewith we synthesized nanoparticles in a range of 6.5\,$\pm$\,0.1 - 176\,$\pm$\,2\,nm, as shown in figure \ref{fig:1}, only variating the quantity of oleylamine. The size distribution is shown in figure \ref{fig:2}. The size reduction may be related to the increased of the ratio surfactant/precursor, but is remarkable the fact that a small change in the added mass of oleylamine can result in a large variation of size. Another interesting result is the shape modification which is attributed to the type of the first coating. There has being previously reported that the type of the functional group of the surfactant determine the final shape of the nanoparticles\cite{nguyen}. The capping reagent can selectively adsorb on some particular facets, but the quantitative result of a mixture has not yet been reported. However, same works explores the growth of the polyhedral form as a function of the surface energy, $\gamma_{hkl}$, where the results agree with ours\cite{plasma}.
The oleic acid with carboxylic group, -COOH, has a selective binding onto the \{111\} crystal facets and oleylamine with a -NH$_2$ group has a weak and unselective binding onto the surface which leads to more spherical particles \cite{smp}. It is also known that at a low heating rate, preferential anisotropic growth along the kinetically most favorable direction, but according to the classical theory, a variation on the surface energy caused by the surfactant is dominant to fluctuations in temperature and saturation.\cite{spc, naturecrystal}.

Thereby, we define a shape factor parameter, $\upsilon$, relating the amount of oleic acid and oleylamine surfactants which bind differently onto crystal facets.
\begin{equation}\label{eq:1}
\upsilon \equiv \frac{n_{\text{oleylamine}}}{n_{\text{oleic acid}}}
\left\{\begin{matrix}
>0&:& \Square / \lozenge \rightarrow \Circle
\\ 
= 0&:& \hexagon
\end{matrix}\right.,
\end{equation}
where $n_{\text{oleylamine}}$ and $n_{\text{oleic acid}}$ are  the number of mols of oleylamine and oleic acid respectively. When $\upsilon=0$ the favorable growth direction is the \{111\} because there is only oleic acid and when added oleylamine, $\upsilon>0$, the shape became a mix of cubes and octahedrons until it becomes spherical with the increase of this shape factor.
A linear relationship between the diameter and the surfactant/precursor ratio is shown in figure \ref{fig:2}. The gray curve is a best linear fitting obtained in a semi-log graph and is represented mathematically in equation \ref{eq:2} as
\begin{equation}\label{eq:2}
d = \frac{d_{wo}}{e^{6.5\frac{\upsilon}{n_{P}}}},
\end{equation}
where $d$ is the diameter of nanoparticles, $d_{\text{wo}}$ is the nanoparticle diameter for a synthesis without oleylamine, 6.5 is a fitted parameter and $n_{P}$ is the number of mol of the precursor.

\begin{figure}
\centering
  \includegraphics[scale=0.45]{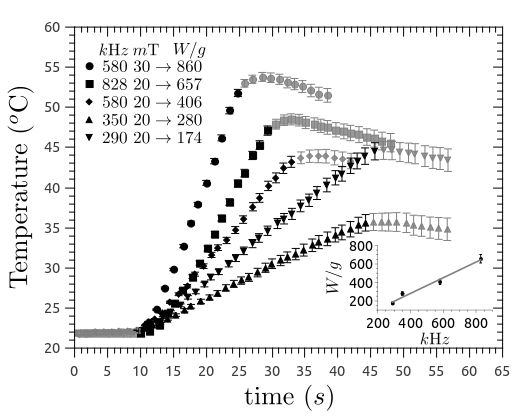}
\caption{Time-dependent temperature curves of nanoparticles with diameter of 29\,nm. The black points indicates when the AC magnetic field was applied. The inset shows the linear dependence between the specific absorption rate and the radio-frequency used.}
\label{fig:3}
\end{figure}

To explore a potential biomedical application, we measure the SAR of the nanoparticles with 29\,nm average size, which is shown in figure \ref{fig:3}.
The results point to a notable specific absorption rate for these magnetite nanoparticles\cite{sar, sar2, sar3} even at low frequencies, becoming appropriate for thermoablation.\cite{aplica}. The Brown relaxation due to the rotation of the particles against viscous forces of the liquid medium is expect for this size, including N\'eel relaxation and the magnetic hysteresis loss which is proportional to the size was enlarged by the double-layer coating\cite{spa2}.  The inset in figure \ref{fig:3} shows a linear dependence between the SAR and the radio-frequency field applied in a 20\,mT magnetic field.

In summary, we synthesized nanoparticles of magnetite with sizes along three orders of magnitude, solely modifying the amount of a unselective binding surfactant which also leads to a shape control. We obtained a relationship between the morphology of the nanoparticles and the ratio surfactant/precursor and we have found an equation which treat this dependence including a shape parameter ($\upsilon$) for this synthesis configuration. Finally, we have explored a possible biomedical application of hyperthermia which proved to be superior to those found in literature.

\begin{acknowledgement}

This work had the financial support of the brazilian agency CAPES (Bolsista da CAPES - Processo n\textordmasculine \ 2263-13-0).

\end{acknowledgement}





\begin{thebibliography}{10}

\bibitem{martian} Thomas-Keprta, K.L.;  Clemett, S.J.;  McKay, D.S.;  Gibson, E.K.;  Wentworth, S.J. (2009)
Geochimica et Cosmochimica Acta	vol. 73 (21) p. 6631-6677 

\bibitem{ink}Rosensweig, R. E. Scientific American (1992), 247 (4), 136

\bibitem{batteries} Ming Zhang;  Danni Lei;  Xiaoming Yin;  Libao Chen;  Qiuhong Li et al. (2010)
Journal of Materials Chemistry	vol. 20 (26) p. 5538-5543 

\bibitem{medicalapplications} Tran, Nhiem;  Webster, Thomas J. (2010) J. Mater. Chem.	vol. 20 (40) p. 8760-8767 

\bibitem{surface1} Nam, Jutaek;  Won, Nayoun;  Bang, Jiwon;  Jin, Ho;  Park, Joonhyuck et al. (2013)
Advanced Drug Delivery Reviews	vol. 65 (5) p. 622-648 

\bibitem{shapeinfluence1} Pal, Sukdeb;  Tak, Yu Kyung;  Song, Joon Myong (2007) Appl. Envir. Microbiol.	vol. 73 (6) p. 1712-1720 

\bibitem{shapeinfluence2} Chithrani, B. Devika;  Chan, Warren C. W. (2007) Nano Lett.	vol. 7 (6) p. 1542-1550 

\bibitem{shapeinfluence4} Huang, Xinglu;  Teng, Xu;  Chen, Dong;  Tang, Fangqiong;  He, Junqi (2010)
Biomaterials	vol. 31 (3) p. 438-448 

\bibitem{shapeinfluence5}  Albanese, Alexandre;  Tang, Peter S;  Chan, Warren C W (2012) Annual review of biomedical engineering	vol. 14 p. 1-16 

\bibitem{shapeinfluence6} Wani, Irshad A.;  Ahmad, Tokeer (2013) Colloids and Surfaces B: Biointerfaces	vol. 101 p. 162-170 

\bibitem{spa1} Lima, E.;  Torres, T. E.;  Rossi, L. M.;  Rechenberg, H. R.;  Berquo, T. S. et al. (2013)

\bibitem{spa2} Goya, G. F.;  Berqu\'o, T. S.;  Fonseca, F. C.;  Morales, M. P. (2003) Journal of Applied Physics	vol. 94 (5) p. 3520-3528 

\bibitem{hypertermia} Branquinho, Luis C.;  Carrião, Marcus S.;  Costa, Anderson S.;  Zufelato, Nicholas;  Sousa, Marcelo H. et al. et al. (2013) Scientific Reports	vol. 3 

\bibitem{Sun2002} Sun, Shouheng;  Zeng, Hao (2002)
J. Am. Chem. Soc.	vol. 124 (28) p. 8204-8205 

\bibitem{dokyoon} Kim, Dokyoon;  Lee, Nohyun;  Park, Mihyun;  Kim, Byung Hyo;  An, Kwangjin et al. (2008)
J. Am. Chem. Soc.	vol. 131 (2) p. 454-455 

\bibitem{lauric} Pradhan, Pallab;  Giri, Jyotsnendu;  Banerjee, Rinti;  Bellare, Jayesh;  Bahadur, Dhirendra (2007) Journal of Magnetism and Magnetic Materials	vol. 311 p. 282-287 

\bibitem{nguyen} Nanotechnology and Nanomaterials "Nanocrystal", book edited by Yoshitake Masuda, ISBN 978-953-307-199-2, Published: June 28, 2011 under CC BY-NC-SA 3.0 license 

\bibitem{plasma} Swaminathan, R.;  Willard, M.A.;  McHenry, M.E. (2006) Acta Materialia	vol. 54 (3) p. 807-816 

\bibitem{smp} Yang, Haitao;  Ogawa, Tomoyuki;  Hasegawa, Daiji;  Takahashi, Migaku (2008) Journal of Applied Physics	vol. 103 (7) p. 07D526 

\bibitem{spc} Franks, G.V.;  Zhang, Ling;  Li, Qin;  Liu, Shaomin;  Ang, Ming et al. (2011)
Advanced Powder Technology	vol. 22 (4) p. 532-536

\bibitem{naturecrystal} Baumgartner, Jens;  Dey, Archan;  Bomans, Paul H. H.;  Le Coadou, Cécile;  Fratzl, Peter et al. (2013) Nature Publishing Group	vol. 12 (4) p. 310-314 

\bibitem{sar} Ma, Ming;  Wu, Ya;  Zhou, Jie;  Sun, Yongkang;  Zhang, Yu et al. (2004)
Journal of Magnetism and Magnetic Materials	vol. 268 

\bibitem{sar2} Deatsch, Alison E.;  Evans, Benjamin A. (2014) Journal of Magnetism and Magnetic Materials	vol. 354 p. 163-172 

\bibitem{sar3} Garaio, E.;  Collantes, J.M.;  Garcia, J.A.;  Plazaola, F.;  Mornet, S. et al. (2013) Journal of Magnetism and Magnetic Materials

\bibitem{aplica} Kumar, Challa S S R;  Mohammad, Faruq (2011) Advanced drug delivery reviews	vol. 63 (9) p. 789-808 
\end{thebibliography}
\end{document}